\documentclass[preprint,12pt]{elsarticleLJW}

\usepackage{amssymb}
\usepackage{amsthm}
\usepackage{graphicx}
\usepackage{float}
\usepackage{mathrsfs}
\usepackage{multirow}
\usepackage{epstopdf}

\begin{document}

\begin{frontmatter}

\title{SARS-Cov-2 RNA Sequence Classification Based on Territory Information}

\author{Jingwei Liu \corref{cor*}}
\ead{liujingwei03@tsignhua.org.cn}
\cortext[cor*]{Corresponding author.}
\address{School of Mathematical Sciences, Beihang University, Beijing, 102206,P.R China}

\begin{abstract}
CovID-19 genetics analysis is critical to determine virus type,virus variant and evaluate vaccines.
In this paper, SARS-Cov-2 RNA sequence analysis relative to region or territory is investigated.
A uniform framework of sequence SVM model with various genetics length from short to long and
mixed-bases is developed by projecting SARS-Cov-2 RNA sequence to different dimensional space,
then scoring it according to the output probability of pre-trained SVM models to explore the
territory or origin information of SARS-Cov-2. Different sample size ratio of training set and test set
 is also discussed in the data analysis. Two SARS-Cov-2 RNA classification tasks are constructed  based on GISAID database, one is for mainland, Hongkong and Taiwan of China,
 and the other is a 6-class classification task (Africa, Asia, Europe, North American, South American\& Central American, Ocean) of 7 continents. For 3-class classification of China, the Top-1 accuracy rate can reach 82.45\% (train 60\%, test=40\%); For 2-class classification of China, the Top-1 accuracy rate can reach 97.35\% (train 80\%, test 20\%); For 6-class classification task of world, when
the ratio of training set and test set is 20\%  : 80\% , the Top-1 accuracy rate can achieve 30.30\%.
And, some Top-N results are also given.

\end{abstract}


\end{frontmatter}

\section{Introduction}
\label{}
 The ongoing COVID-19 pandemic has led disaster to human being all over the world. From first identified in Wuhan, China late December 2020 to 8 January 2021, more than 88.7 million persons are infected, and more than 1.91 million people are dead attributed to severe acute respiratory syndrome coronavirus 2 (SARS-Cov-2). The origin of CovID-19 is still a concerned issue, thought there is a scientific consensus that it has a natural origin. [1,2,3,4,5,6]

CovID-19 pandemic poses challenges to the state-of-the-art techniques of genome analysis combined with artificial intelligent(AI) and machine learning (ML), which is employed in CovID-19 pandemic for virus discovery,virus variant, virus evolvement, genetic mutation, symptom diagnosis, conditions mostly affecting the spread and development of conventional drugs and vaccines.[7-26]

It is well-known that SARS-Cov-2 is various among different hosts. This paper aims to investigate
the  diversity of SARS-Cov-2 among different people related to different region and territory, and reports our tracking research on the SARS-Cov-2 genetic sequences in CovID-19 pandemic by designing two pattern recognition tasks chronologically following submission time in GISAID. The first one is of SARS-Cov-2 from three regions of China, mainland, Hongkong and Taiwan as of 2 June,2020, attempting to explore the origin of SARS-Cov-2 in China in the view of region information. Our result show that the SARS-Cov-2 RNA sequences of Hongkong are distinctly discriminated from those of mainland and Taiwan. The second one is a 6-class pattern recognition task designed from 7 continents worldwide as of 11 July,2020, to investigate origin and diversity of SARS-Cov-2, which is helpful to estimate the vaccine effectiveness for CovID-19 pandemic.

The paper is organized as follows: Section 2 describes our sequence SVM model with Top-N, Section 3 introduces experimental database , Section 4 presents the experimental results, Section 5 summarizes the research.

\section{Method}
\label{}
\subsection*{Sequence Multi-class SVM with Top-N}

Support Vector Machine (SVM) is an efficient supervised statistical learning and machine learning method for classification analysis. The multi-class SVM is defined on binary SVM with one-versus-one max-wins voting strategy or one-versus-all winner-takes-all way. The Top-N method is also called ranking technique or N-best method, it is widely applied in pattern recognition and recommendation system. A sequence SVM model with Top-N is developed for microbial marker clades gene sequence classification in [27].Later, the motivation also appears in [28]. 

The same  sequence SVM model with Top-N is employed in this paper to dig out origin and territory information for SARS-Cov-2. The novel treatment here is that the mixed-base is also involved in genetics analysis, it is treated as a noise of genome sequence, however there should have information in the site of SARS-Cov-2 RNA sequence though the exact base is not measured.

To strengthen the confidence of experiment results, the standard LibSVM  
\{ http://www.csie.ntu.edu.tw/~cjlin/libsvm/ \} is utilized in training stage, and the parameter range of C-SVM  with RBF kernel is  $C\times \gamma $ $\in$ \{0.0625, 0.125, 0.25, 0.5, 1, 2, 3, 4 \} $ \times $\{0.0625, 0.125, 0.25, 0.5, 1, 2, 3, 4 \}. Then, the Top-N stage is complemented with C++ platform with LibSVM.
The best classification accuracy rates of test set (independent from training set ) in total 64 parameters are reported below.

\section{Dataset}
\label{}

All of the SARS-Cov-2 genome sequences are download from the GISAID database (https://www.gisaid.org/) according to the submission time.
Data-I is about 3 regions, mainland, Hongkong and Taiwan, of China as of 2 June, 2020, totally 752 genome sequences with length from 107 to 30057, where 29 RNA sequences lengths range from 107 to 835, and the others' lengths are from 1065 to 30057. The average length is 25360. And the sample numbers of mainland, Hongkong and Taiwan are 508,159,85 respectively.

Data-II is a 6-class (Africa, Asia, Europe, North American, South American\& Central American, Ocean) experimental dataset from 7 continents. The total genome sequence number is 9904 with length varying from 87 to 30643, where 84 RNA sequences' lengths range from 87 to 900, and the others' lengths are from 1065 to 30643. The average length is 29319. Considering the big-data size of  SARS-Cov-2 genomic sequence in GISAID, the selection rule is that sample number of each class is no more than 2000, and as of 11 July,2020, the sample number of the above 6-class is 845,2000,2000,2000,2000,1059 respectively.

The pre-process of pattern recognition is as follows. For Data-I, there are pure genetic sequence of acid base \{A, C, T, G\} , but there are mixed-base or ambiguous base in the genome sequences of Data-II. All of the acid bases sequences involved in the experiments are downloaded chronologically with submission time of GISAID without abandon.
In pattern recognition task. The acid base \{A, C, T, G\} is mapped into \{1,2,3,4\} respectively and the mixed-base is mapped as the average value of pure acid base involved above.

The genetics sequences are split into training set and test set according to given split percent \{0.2, 0.4, 0.6, 0.8\}. then all the sequences in training set and testing set are mapped into k--dimension sample space, where k-dimension is also treated as k-mer size. To investigate the dimension of space for discrimination, the k-mer sizes of \{10, 20, 30, 40, 50, 60, 70,80 \} are discussed in the experiments, and the overlaps of \{0, 25\%,  50\%, 75\% \} percent of adjacent k-mer length vectors in original sequence are also discussed. The pattern recognition rates of test set under above pre-process are briefly denoted as ``S\{split percent\}'', ``O=overlap'', which means the accuracy rate is obtained using training model with given split percent set to predict the rest test set with (1-split percent) of whole database.

After pre-process, all the data sets are utilized in training and classification phases separately according to their respective tasks.

\section{Experimental Result}
\label{}

\subsection{3-class SARS-Cov-2 classification of China }
\label{}

In this experiment, SARS-Cov-2 genome sequences of China are categorized into a 3-class pattern classification problem related to mainland, Hongkong and Taiwan respectively. The sequence SVM accuracy rates of Top-1 and Top-2 are listed in Figure 1 and Figure 2 respectively. The experimental results show that the best accurate rate can be reached under parameter adjustment, even in case of S0.2 which is training set (20\%) and test set (80\%). And, the accuracy rate S0.8 takes no absolute advantage of S0.2 in Top-1 case. The overall  Top-2 classification rate is higher than that of Top-1, It indicates that there are some samples with high similarity, after all, the three regions are within China, at least almost same race though belonging to different ethnic groups.

\begin{figure}[!htbp]
\begin{center}
\begin{minipage}{\textwidth}
\includegraphics[width=\textwidth]{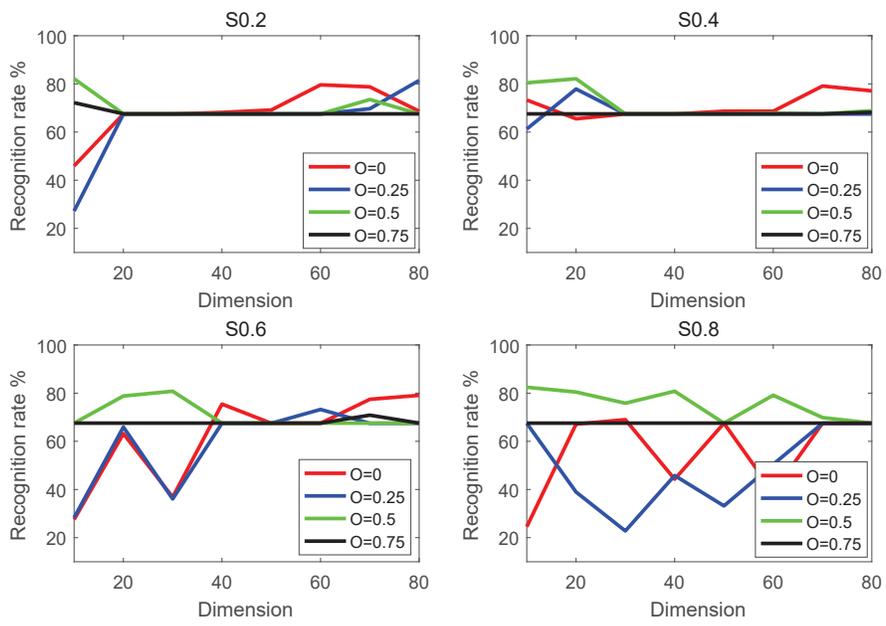}
\caption{3-class  Top-1 accuracy rate of mainland,Hongkong and Taiwan. }
\end{minipage}
\end{center}
\end{figure}

\begin{figure}[!htbp]
\begin{center}
\begin{minipage}{\textwidth}
\includegraphics[width=\textwidth]{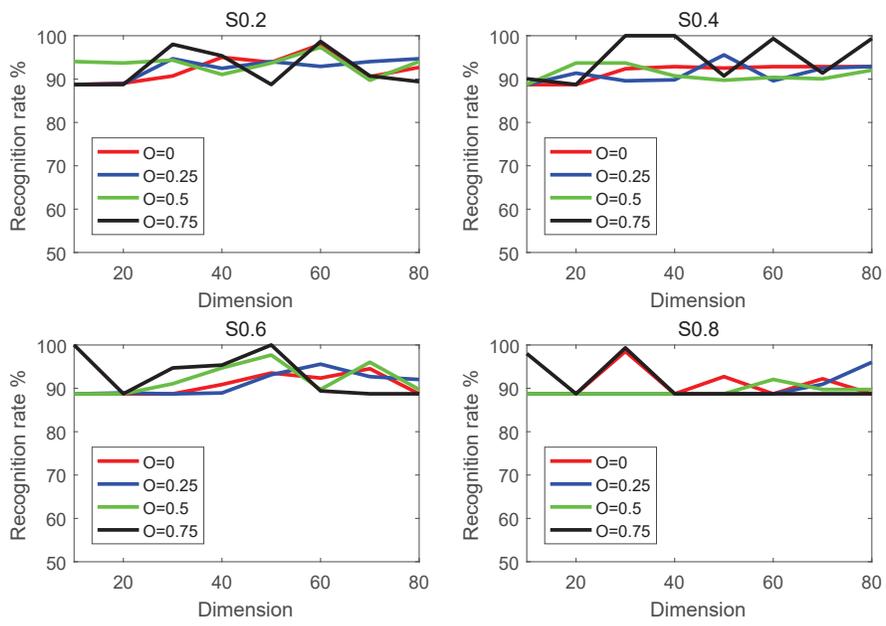}
\caption{3-class Top-2 accuracy rate of mainland,Hongkong and Taiwan. }
\end{minipage}
\end{center}
\end{figure}

\subsection{2-class SARS-Cov-2 classification of China}
\label{}

To explore the sample structure of SARS-Cov-2 of China, three 2-class pattern classification experiments are designed, that is any two of the above regions are combined into one class, and the other are treated as one class. The Top-1 classification rates are shown in Figure 3, Figure 4, Figure 5. The accuracy rate of Hongkong and mainland \& Taiwan get the best result (Figure 4), it can reach the conclusion that the SARS-Cov-2 virus of Hongkong is different from that of  mainland \& Taiwan.
Comparing the results of 3-class and 2-class of  SARS-Cov-2 of China, it also indicates that with the increase of class number, the accuracy rate will decrease. And, in the case of 2-class classification (Figure 3, Figure 4, Figure 5),  the accurate rate will obtain high value with the increase of split percent. But, this advantage no longer holds in 3-class recognition (Figure 1,Figure 3,Figure 4,Figure 5).

\begin{figure}[!htbp]
\begin{center}
\begin{minipage}{\textwidth}
\includegraphics[width=\textwidth]{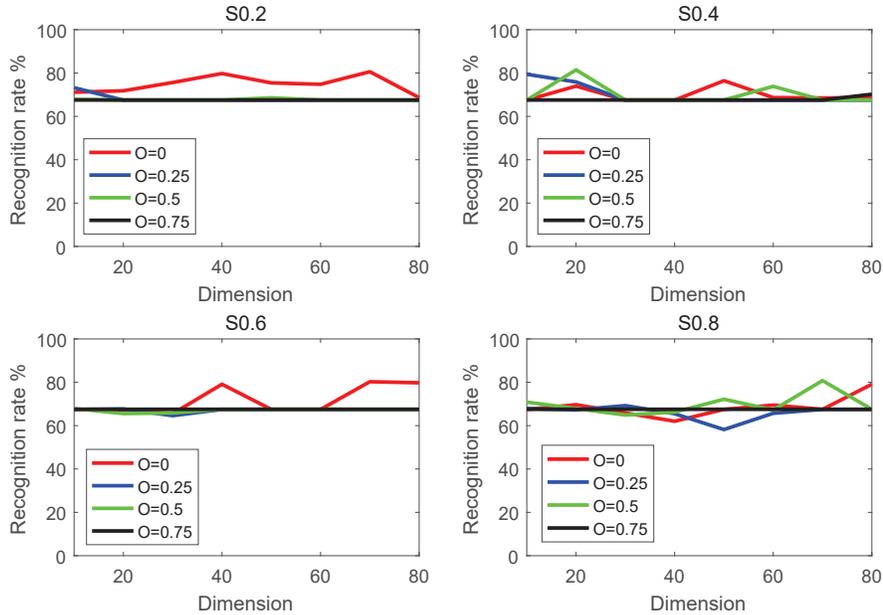}
\caption{2-class classification of mainland vs. Hongkong and Taiwan.}
\end{minipage}
\end{center}
\end{figure}

\begin{figure}[!htbp]
\begin{center}
\begin{minipage}{\textwidth}
\includegraphics[width=\textwidth]{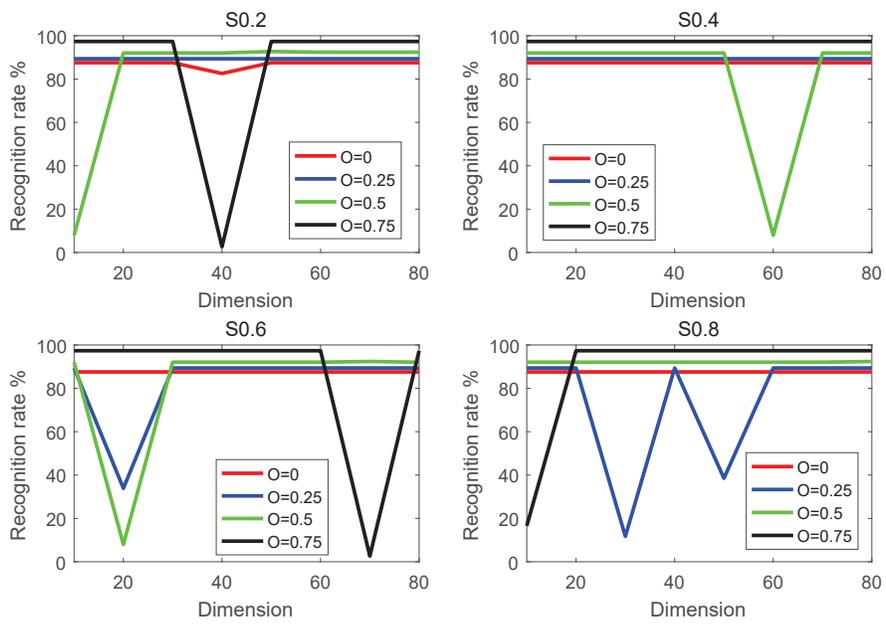}
\caption{2-class classification of Hongkong vs. mainland and Taiwan.}
\end{minipage}
\end{center}
\end{figure}

\begin{figure}[!htbp]
\begin{center}
\begin{minipage}{\textwidth}
\includegraphics[width=\textwidth]{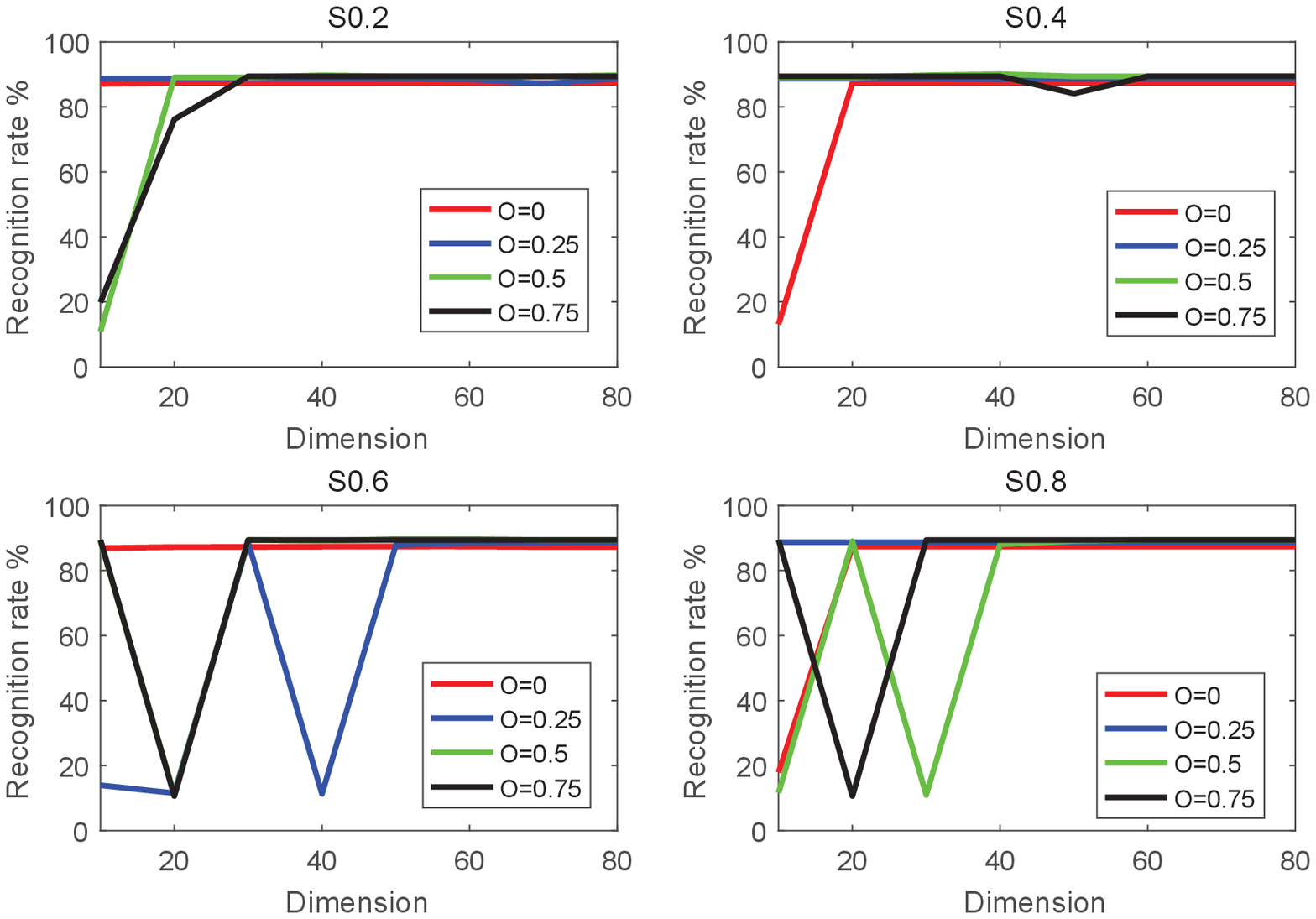}
\caption{2-class classification of Taiwan  vs. mainland and Hongkong.}
\end{minipage}
\end{center}
\end{figure}

\subsection{6-class SARS-Cov-2 classification worldwide}
\label{}

As illustrated above, the accuracy rate with more training set to predict the less test set may not take advantage with the increase of class number, the causes may be explained from two aspects, the first one is
that it is determined by the  characteristics of SARS-Cov-2 RNA sequence, the other explanation is that all the
SARS-Cov-2 genetic sequences are of human beings. However, the collection of SARS-Cov-2 of CovID-19 pandemic is from less to more, and the development of vaccine is also applied limited SARS-Cov-2 virus samples and infected persons. Using less SARS-Cov-2 genetic sequences to predict or classify more SARS-Cov-2 genetic sequences still has significance in the special case.

For the purpose of  revealing diversity of human being SARS-Cov-2 and explore the origin of SARS-Cov-2,
limited to the computation time of big data, and the fact that vaccine development only involved a bit samples, the accurate rates with split rate 0.2 , and overlap 0 and 0.25 are reported in Figure 6 and Figure 7.
The above designment coincides the prediction and classification of quick outbreak of CovID-19.
The classification rate with training set 20\% to predict test set 80\% show that there are about 30.30\% genetic sequences of SARS-Cov-2 which are related to region or territory, which provides a data analytical consideration for estimation of vaccine.

\begin{figure}[!htbp]
\begin{center}
\begin{minipage}{\textwidth}
\includegraphics[width=\textwidth]{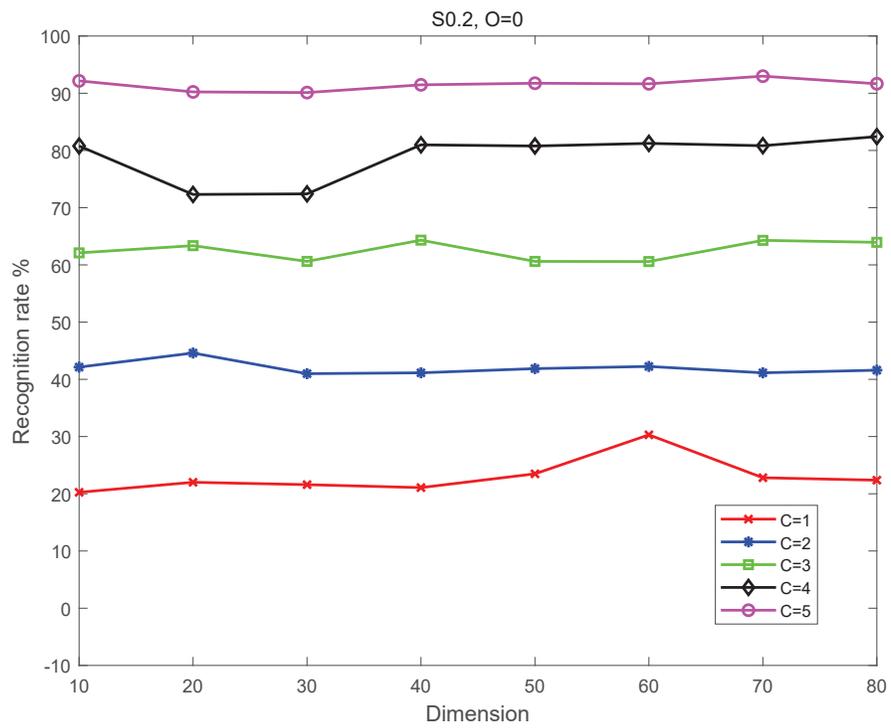}
\caption{6-class different Top-N classification rates with overlap 0 and split rate 0.2 , where C denotes number of candidate in Top-N.}
\end{minipage}
\end{center}
\end{figure}

\begin{figure}[!htbp]
\begin{center}
\begin{minipage}{\textwidth}
\includegraphics[width=\textwidth]{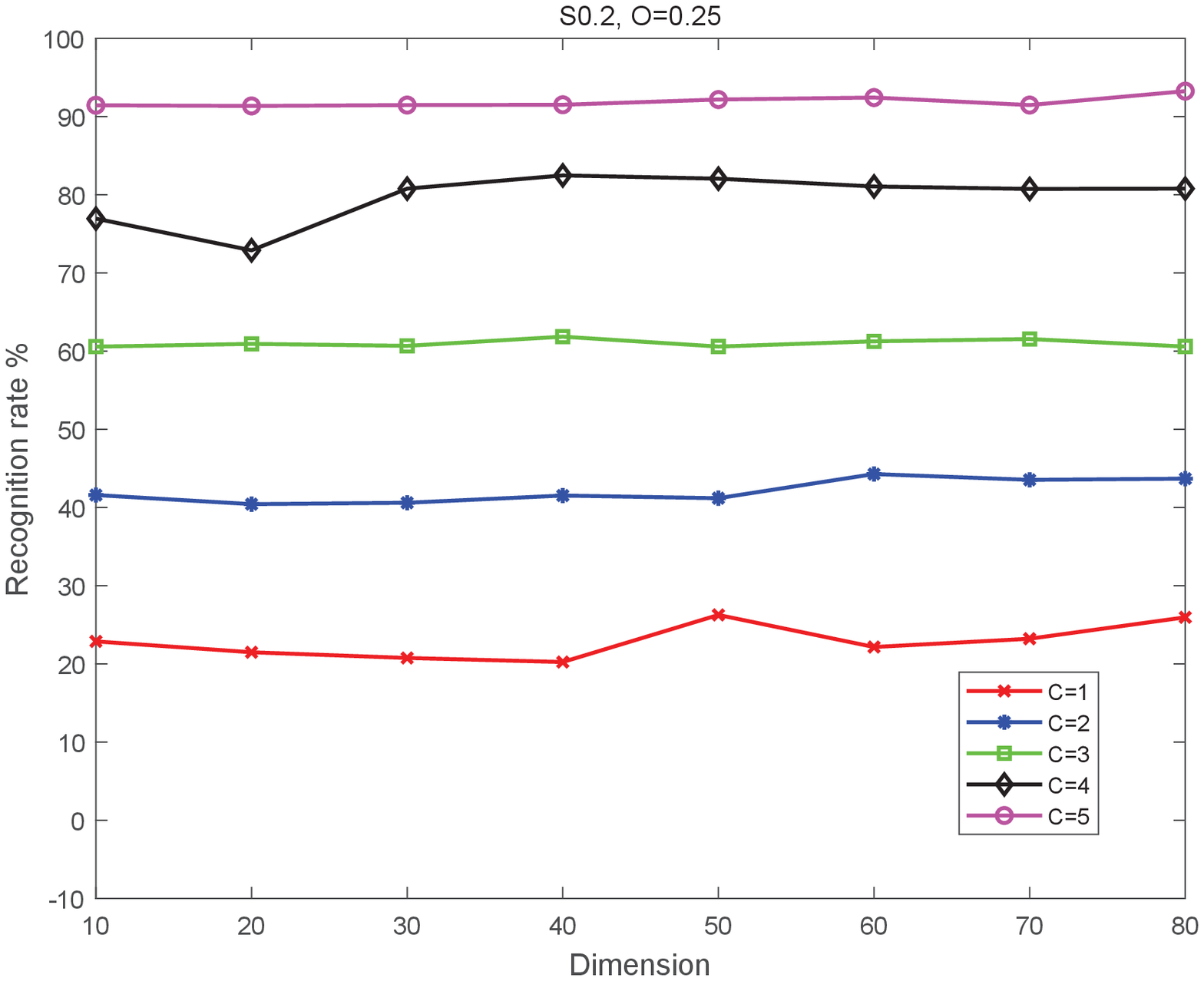}
\caption{6-class different Top-N classification rates with overlap 0.25 and split rate 0.2, where C denotes number of candidate in Top-N.}
\end{minipage}
\end{center}
\end{figure}

The accuracy rates with different overlap in big data SARS-Cov-2 modelling are almost similar. And, with the
increase of candidate number of Top-N, the accuracy rates tends to almost uniformly increase with Top-N candidates in both cases of ``O=0'' and ``O=0.25'' (Figure 6, Fingure7 ). It also illustrates that beside about
30\% discriminated  SARS-Cov-2 RNA sequences , the rest SARS-Cov-2 RNA sequences are still highly confused and show almost well-proportioned distribution in view of Top-N.

\section{Conclusion}
\label{}

SARS-Cov-2 RNA sequences analysis related to region and territory will provide new deep insights into the behavior of CovID-19. Using less train set to predict more test set help us to estimate the trend of CovID-19 pandemic, variant capability of SARS-Cov-2, and application range of vaccines.
Our experimental results demonstrate that, with the SARS-Cov-2 before 6 June,2020, the SARS-Cov-2 genome sequences are high discriminated the SARS-Cov-2 genome sequences of mainland \& Taiwan. And, among the worldwide SARS-Cov-2 virus, there are 30\% RNA virus which are apparently differentiated in mathematical space.

\end{document}